\def\year{2020}\relax
\documentclass[letterpaper]{article} 
\usepackage{aaai20}  
\usepackage{times}  
\usepackage{helvet} 
\usepackage{courier}  
\usepackage[hyphens]{url}  
\usepackage{graphicx} 
\usepackage{graphicx}  
\usepackage{newtxmath} 
\usepackage{booktabs}
\usepackage{threeparttable}
\newtheorem{mydef}{Definition}
\usepackage{subcaption}
\usepackage[ruled,linesnumbered]{algorithm2e}
\newcommand{\citet}[1]{{\citeauthor{#1} \shortcite{#1}}}
\title{Author Name Disambiguation on Heterogeneous Information Network with Adversarial Representation Learning}

\author{
Haiwen Wang,\textsuperscript{\rm 1} 
Ruijie Wang,\textsuperscript{\rm 2} 
Chuan Wen,\textsuperscript{\rm 1} 
Shuhao Li,\textsuperscript{\rm 1} \\
\bf \Large 
Yuting Jia,\textsuperscript{\rm 1} 
Weinan Zhang,\textsuperscript{\rm 1} 
Xinbing Wang\textsuperscript{\rm 1}  \\
\textsuperscript{\rm 1}Shanghai Jiao Tong University, Shanghai, China\\
\textsuperscript{\rm 2}University of Illinois at Urbana-Champaign, Urbana, IL, USA\\
\{wanghaiwen, AlvinWen, cherish\_a, hnxxjyt, wnzhang, xwang8\}@sjtu.edu.cn, 
ruijiew2@illinois.edu}

\begin{document}
\maketitle

\begin{abstract}

Author name ambiguity causes inadequacy and inconvenience in academic information retrieval, which raises the necessity of author name disambiguation (AND). 
Existing AND methods can be divided into two categories: the models focusing on content information to distinguish whether two papers are written by the same author, the models focusing on relation information to represent information as edges on the network and to quantify the similarity among papers. 
However, the former requires adequate labeled samples and informative negative samples, and are also ineffective in measuring the high-order connections among papers,
while the latter needs complicated feature engineering or supervision to construct the network. 
We propose a novel generative adversarial framework to grow the two categories of models together:
(i) the discriminative module distinguishes whether two papers are from the same author, and (ii)
the generative module selects possibly homogeneous papers directly from the heterogeneous information network, which eliminates the complicated feature engineering.
In such a way, the discriminative module guides the generative module to select homogeneous papers, and the generative module generates high-quality negative samples to train the discriminative module to make it aware of high-order connections among papers.
Furthermore, 
a self-training strategy for the discriminative module and a random walk based generating algorithm are designed to make the training stable and efficient.
Extensive experiments on two real-world AND benchmarks demonstrate that our model provides significant performance improvement over the state-of-the-art  methods.

\end{abstract}

\section{Introduction}

A person name is used to identify a certain individual. 
However, different people may have the same or the similar name in the real world, which is referred to as the name ambiguity.  
For example, \texttt{Michael J} can remind people of the US basketball player, the King of Pop or the machine learning professor from UC Berkeley. 
Name ambiguity causes inadequacy and inconvenience in information retrieval. 
With the rapid development of the scholar community, academic information in digital libraries becomes increasingly tremendous.
However, names appearing in the digital papers or the webpages also suffer from the ambiguity issues, which means that the author name cannot be used to reliably identify all scholarly authors.
The inadequacy of author name ambiguity becomes evident in many practical scenarios, e.g., scholar searching, influence evaluating and mentor recommendation, which raises the necessity of author name disambiguation \cite{smalheiser2009author}.

Author name disambiguation is to split the papers under the same name into several homogeneous groups, 
which has attracted substantial attention from information retrieval and data mining communities. 
Most existing methods solve this problem in a two-stage framework: 
(i) quantify the similarity among papers; 
(ii) cluster papers into homogeneous groups. 
Hierarchical clustering algorithm works well for the second part, while the first part remains largely unsolved.
To quantify the similarity among papers, content information and relation information are used. 
The former includes title, abstract, introduction and keywords etc.   
Methods focusing on the content information \cite{han:2014,Huang:2006,Yoshida:2010} usually leverage supervised learning algorithms to learn the pairwise similarity functions. 
However, they solve the problem in a local way, which means that they cannot measure the high-order connections among papers.
Methods focusing on relation information \cite{Kanani:2007,Bekkerman:2005} usually solve the problem on the bibliographic network, where the relation information is represented as edges on the network.
They account that papers connected in the network are likely to be written by the same author.
Thus constructing the network becomes the critical part of these methods, e.g., paper network \cite{JieTang}, paper-author network \cite{Anonymized}.
However, either complicated feature engineering or the supervision \cite{JieTang} is required. 

The two categories of methods are like the two sides of the same coin. 
The first introduces supervision but cannot process high-order connections, while the second models the high-order connections but requires the supervision.
An intuitive idea is to combine them together to build a unified model which can eliminate the requirement of labeled samples and complicated feature engineering to some extent. 
Inspired by generative adversarial networks \cite{GAN}, we may combine the two categories in an adversarial way.
In this paper, we propose a unified framework with discriminative module and generative module. 
The discriminative module directly distinguishes whether two papers are written by the same author based on feature vectors. 
This module is learned in a self-training way, and it requires negative samples generated by the generative module. 
The generative module works on the heterogeneous information network and selects papers viewed as the homogeneous pairs.

In this framework, the discriminative module can guide the exploration of the generative module to select homogeneous papers on the raw network. 
And the generative module can generate high-quality samples with high-order connections for the discriminative module, which can make it aware of the topology of the networks. 
We verify the performance of the proposed model on two benchmark datasets. The results demonstrate the significant superiority of our proposed method over the state-of-the-art author name disambiguation solutions.

In sum, the contributions of this paper are three-fold.
\begin{itemize}
\item We comprehensively take the content information and relation information into consideration by constructing the heterogeneous information network which eliminates the requirement for complicated feature engineering.
\item We design a unified framework combining a discriminative module and a generative module based on the heterogeneous information network for author name disambiguation task.
Experimental results on two real-world datasets verify the advantages of our method over state-of-the-arts.
\item To support AND research, we construct a sufficiently large benchmark dataset consisting of 17,816 authors and 130,655 papers. Compared with the existing benchmark datasets, it is the largest AND dataset with rich content information and relation information.
\end{itemize}

\section{Related Work}

\vspace{5pt} \noindent \textbf{Author Name Disambiguation.}
To measure the similarity among papers, the existing methods can be divided into two categories according to the information they focus on.
The first are based on the content information \cite{han:2014,Huang:2006,Block,Yoshida:2010}, which usually solve the problem in a discriminative way.
These methods calculate the content similarity with the help of TF-IDF, exact-matching, and etc.
Then they train supervised models by the labeled samples.
\citet{han:2014} present supervised disambiguation methods based on SVM and Na\"\i ve Bayes.
\citeauthor{Huang:2006} use blocking technique to group candidate papers sharing similar names together. Then it learns distance among papers by SVM.
\citet{Block} use a classifier to learn pairwise similarity and perform semi-supervised hierarchical clustering. 
The problem of these models, except for the requirement for the labeled samples,  is that they only take the pairwise similarity into consideration. 
They ignore the high-order connections.
To address the problem, some methods focus on the relation information from the network \cite{Anonymized,Hermansson:2013} in a generative way.
\citet{Tang:2012} employ Hidden Markov Random Fields to model node features and edge features in a unified probabilistic framework.
\citet{Anonymized} firstly apply network representation learning algorithm into this task on the three constructed graphs based on document similarity and co-authorship.
\citet{JieTang} construct paper networks, where the weights of edges are decided by a supervised model based on the sharing information between two papers. 
These models account that papers connected in the network are likely to be written by the same author. Thus they take the high-order connections into consideration.
And \citet{JieTang} actually transform the academic network into a homogeneous paper network after a complicated feature engineering.
With the help of network representation learning algorithm, we expect a unified model which eliminates the requirement of labeled samples and complicated feature engineering to process the abundant relation information in the network.

\noindent \textbf{Network Representation Learning.}
Network representation learning (NRL), also known as network embedding, aims to learn a low-dimensional representation of each node.
Deepwalk \cite{DeepWalk} first uses random walk and skip gram algorithm inspired by word2vec \cite{word2vec1,wor2vec2} to learn vertex representations.
Node2Vec \cite{Node2vec} applies BFS and DFS search to random walk in order to extract better topology information.
LINE \cite{LINE} tries to preserve both of first-order and second-order network structures.
Some literature explores NRL on heterogeneous networks \cite{PTE,Metapath2Vec}.
However, existing algorithms are designed to preserve the topology information of the network in an unsupervised way. 
We implement it by the reward from the discriminative model in an adversarial framework.

\noindent \textbf{Generative Adversarial Networks.}
Recently, generative adversarial nets (GAN) \cite{GAN} has attracted a great deal of attention. Original purpose of GAN is to generate data from the underlying true distribution, e.g., image \cite{image}, sequence \cite{seq}, dialogue \cite{Dialogue}. Some following literature modifies the framework for the purpose of the adversarial training.
IRGAN \cite{IRGAN} unifies generative model and discriminative model in information retrieval, where the discriminative model provides guidance to the generative model, 
and the generative model generates difficult examples for the discriminative model.
GraphGAN \cite{GraphGAN} combines a designed generative model called \texttt{Graph Softmax} which tries to approximate the
underlying true connectivity distribution and a discriminative model which predicts whether the edge exists between two nodes.
KBGAN \cite{KBGAN} implements the similar motivation in knowledge embedding task, which uses one compositional model as a generator to generate high-quality negative samples for the discriminative model.

\begin{figure}[t]
\centering
\includegraphics[width=0.4\textwidth]{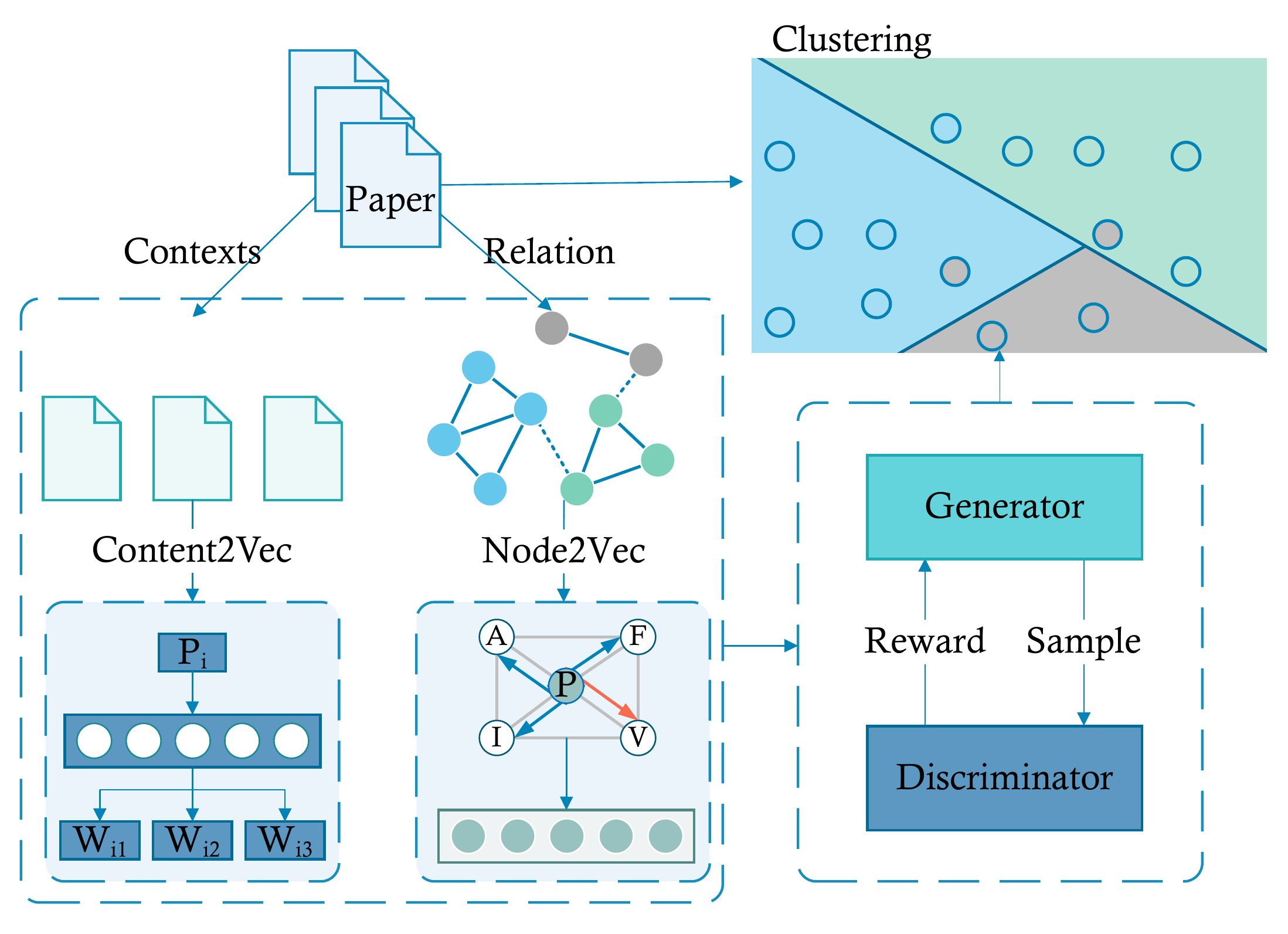}
\caption{An overview of the proposed framework.}
\label{fig:framework}
\end{figure}

\section{Preliminaries}

\subsection{Problem Formulation}
Given an author name reference $a$, let $P^a =\{p^a_1, \dots, p^a_N\}$ be a set of $N$ papers written by the authors with name $a$. 
Each paper $p_i^a \in P^a$ has the content feature set $I_i^a$ including title, abstract, publish date, and etc. 
And it has the relation feature set $R_i^a$, which contains the relation of paper $p^a_i$ to the entities in the academic domain including co-author, institute, field of study and venue.
Given this, we define the problem of author name disambiguation as follows.  
\begin{mydef}
Author Name Disambiguation. The task is to find a function $\Phi$ to partition $P^a$ into a set of disjoint clusters based on the content and relation feature sets $I^a$, $R^a$, i.e.,
$$\Phi(P^a|I^a, R^a) \to C^a, where \ C^a =\{C^a_1, C^a_2, \dots, C^a_k\},$$
where $C^a_i$ means the homogeneous paper subset written by the $i$th author named $a$.
\vspace*{2pt}
\end{mydef}
We omit the superscript $a$ in the following description if there is no ambiguity.
\begin{mydef}
Paper homogeneity. For the convenience of discussion, we define two papers are homogeneous if and only if they are written by the same author.
\end{mydef}
Furthermore, let $y_{ij}$ denote the homogeneity of papers $p_i$ and $p_j$, where $y_{ij}=1$ if $p_i$ and $p_j$ are homogeneous, and $y_{ij}=0$ otherwise.
We denote the generated negative samples as $S_{generated}$ consisting of $(p_i, p_j, y_{ij}=0)$, and the pseudo-positive samples for self-training as $S_{pseudo}$ consisting of $(p_i, p_j, y_{ij}=1)$.

\subsection{Heterogeneous Information Network}
We solve this task with the help of academic heterogeneous information network (HIN), thus content information and relation information can be efficiently processed.
We define the HIN as follows:
\begin{mydef}
Heterogeneous Information Network. The HIN under name reference $a$ is defined as $G = (V, R, I)$ where $V$ is the vertex set including paper, co-author, field of study, institute and venue respectively, and $R=\bigcup _{T = V \backslash P} P \times T $ is the relation set representing the relations among papers and other classes of vertecs, and $I$ is the content information of each $p\in P$.
\end{mydef}

\section{Framework}

The proposed framework is shown in Figure~\ref{fig:framework}. 
In order to represent the information of the heterogeneous information network, we first embed content information and relation information into low-dimension representation space, where two papers are close in the feature space if they are similar. 
Then to integrate the content information and the relation information, and to select homogeneous papers in an adversarial way, we employ a generative adversarial module. 
The generative module aims to explore possible homogeneous papers from the heterogeneous information network, while the discriminative module tries to distinguish the generated negative papers and pseudo-positive papers. 
In such a way, the reward from the discriminative module guides the exploration for the generative module to select homogeneous papers. Moreover, the high-quality papers generated with high-order connections can make the discriminative module aware of the topology of the network.
\begin{figure}[t]
\centering
\includegraphics[width=0.46\textwidth]{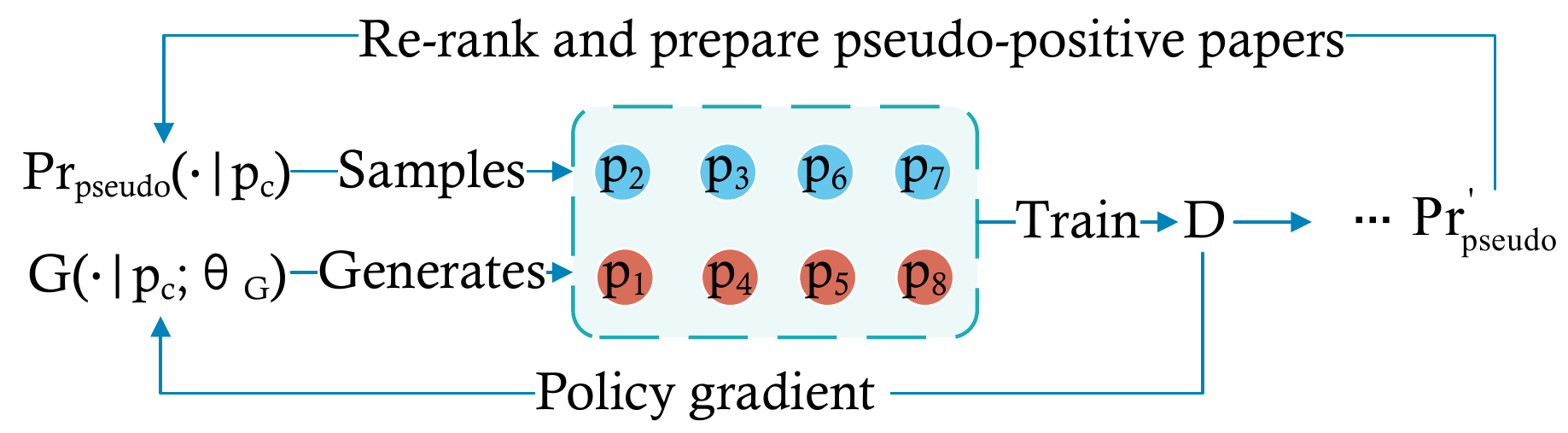}
\caption{The framework of generative adversarial module.}
\label{fig:GAN_framework}
\end{figure}
\subsection{Representation Learning Module}
\subsubsection{Content representation} \label{sec:content}
Papers written by different authors have various topics and literary styles. 
We extract those content features by integrating Doc2vec module \cite{Doc2vec} into our framework. This module learns a low dimension vector $\vec{u}_i \in \mathbb{R}^k$ to represent the information from the content feature set $I_i$ of paper $p_i$. The module updates the parameters by maximizing the log probability of the content sequence, i.e.,
\begin{equation}
\centering
\theta_{\vec{u}} = \mathop{\arg\max}_{\theta} \sum \log \text{Pr}(w_{-b}:w_{b}|p_i; \theta),
\label{eq:doc2vec}
\end{equation}
where $w$ is a word in $I_i$ of $p_i$, $b$ is the window size of the word sequence. 
After optimizing this objective, we can obtain the content representation $\vec{u}_i$ of each paper $p_i$.

\subsubsection{Relation representation} \label{sec:relation}
The topology of HIN we described integrates the relation features of papers. 
Papers having relation features in common are connected in the HIN.
Consequently, We can represent the relation features by preserving the connectivity information of the HIN. 
We use node2vec \cite{Node2vec} to represent these features by $v_i \in \mathbb{R}^k$, where papers are close in the feature space if they have similar relation information. 

\subsection{Generative Adversarial Module}

The core part of our model is shown in Figure \ref{fig:GAN_framework}, integrating content information and relation information of the papers in an adversarial way.
A self-training strategy is added to our discriminative model, which uses top-relevant papers as positive samples iteratively. 
And to make the generative module aware of relation information, following \cite{GraphGAN}, we design a random walk based generating strategy.
Given a paper $p_k$, we design two modules as follows: 

\vspace*{1pt}
\textbf{Discriminative module.} $D(p, p_k; \theta_D)$, which outputs a scalar possibility of whether two papers are written by the same author, i.e., $\text{Pr}(y_{ik}=1|p_i,p_k)$, where $p_i \in P$.

\vspace*{2pt}
\textbf{Generative module.} $G(p|p_k; \theta_G)$, which learns to select the possible homogeneous papers under the guidance of reward. It will iteratively approximate the true underlying homogeneity distribution $\text{Pr}_{true}(p|p_k)$.

Two modules are combined by playing a minimax game: 
the generative module will try to choose the papers possibly written by the same author as the given paper $p_k$, and therefore can fool the discriminative module; 
the discriminative module will distinguish between the selected papers and the ground truth papers.
Formally, generative module $G$ and discriminative module $D$ are playing the following two-player minimax game with value function $V(G, D)$:
\begin{equation}
\begin{split}
\small
\min_{\theta_G}  \max_{\theta_D} & V(G, D) = \sum_{p_k \in P} (\mathbb{E}_{p \sim \text{Pr}_{true}(\cdot|p_k)}[\log D(p, p_k; \theta_D)] \\
&+ \mathbb{E}_{p \sim G(\cdot|p_k;\theta_G)} [\log(1-D(p, p_k; \theta_D)]).
\label{eq:minimax}
\end{split}
\end{equation}

The trainable parameters are the representation of all papers. They are learned by alternately minimizing and maximizing the value function in Eq.~ \eqref{eq:minimax} until the training procedure converges. 
\begin{figure}[t]
\centering
\includegraphics[width=0.48\textwidth]{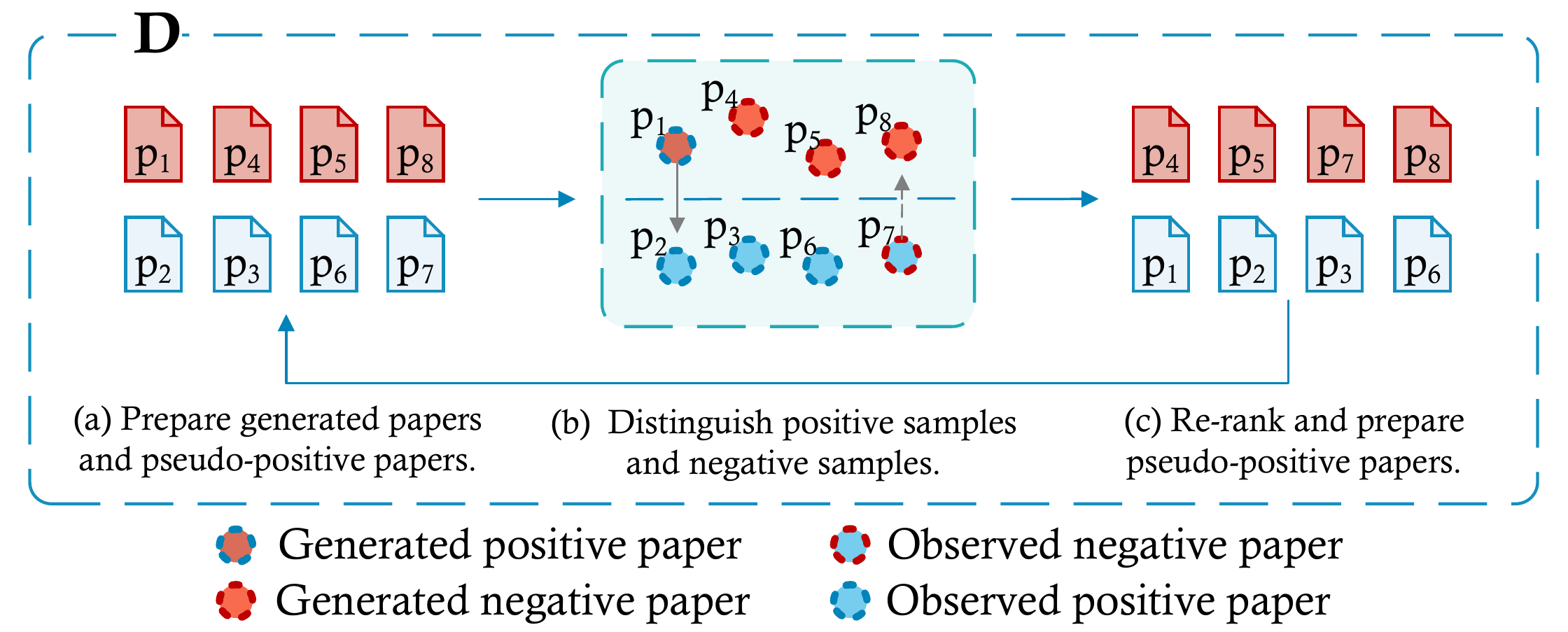}
\caption{Self-training strategy of the discriminative module.}
\label{fig:D_framework}
\end{figure}
\subsubsection{Implementation of Discriminative Module}
Given a paper pair ($p, p_k$), 
we employ a two-layer neural network as our discriminative module to integrate $\vec{u}$ and $\vec{v}$ together:
\begin{equation}
\centering
\vec{d}_{p_i}=\delta(\vec{W}_1^T\delta(\vec{W}_0^T[\vec{u}_i,\vec{v}_i]+\vec{b}_0)+\vec{b}_1),
\label{eq:d}
\end{equation}
\vspace{-5pt}
\begin{equation}
\centering
D(p, p_k)=\text{sigmoid}(\vec{d}_{p}^T\vec{d}_{p_k}),
\label{eq:Dobjective}
\end{equation}
where $\delta(\cdot)$ is non-linear activation function, $\vec{d}_{p_i}$ is the representation vector of paper $p_i$ for $D$, and $\theta_D$ is the union of all $\vec{d}$.
According to Eq.~\eqref{eq:Dobjective}, the content information and relation information can be integrated simultaneously in $\vec{d}$.

To eliminate the requirement for labeling process, we apply the idea of self-training \cite{self1,self2} to select positive samples. Before each $D$ iteration, we select the top possibility papers based on the present results. The selected papers are viewed as a pseudo-positive sample set in the next training process until another selection is performed. The training process of the discriminative module is shown in Figure~\ref{fig:D_framework}.
We update $\vec{d}$ by ascending the gradient concerning the pseudo-positive samples and the generated negative samples:

\begin{equation}
\small
\centering
\nabla_{\theta_D} = \begin{cases} \nabla_{\theta_D}(\log(D(p, p_k)), & \mbox{if } (p, p_k, 1) \in \mbox{ $S_{pseudo}$}; \\ 
\nabla_{\theta_D}(\log(1-D(p, p_k)), & \mbox{ if }  (p, p_k, 0) \in\mbox{ $S_{generated}$}.\\ 
\end{cases}
\label{eq:Dgradient}
\end{equation}

\begin{figure}[t]
\centering
\includegraphics[width=0.48\textwidth]{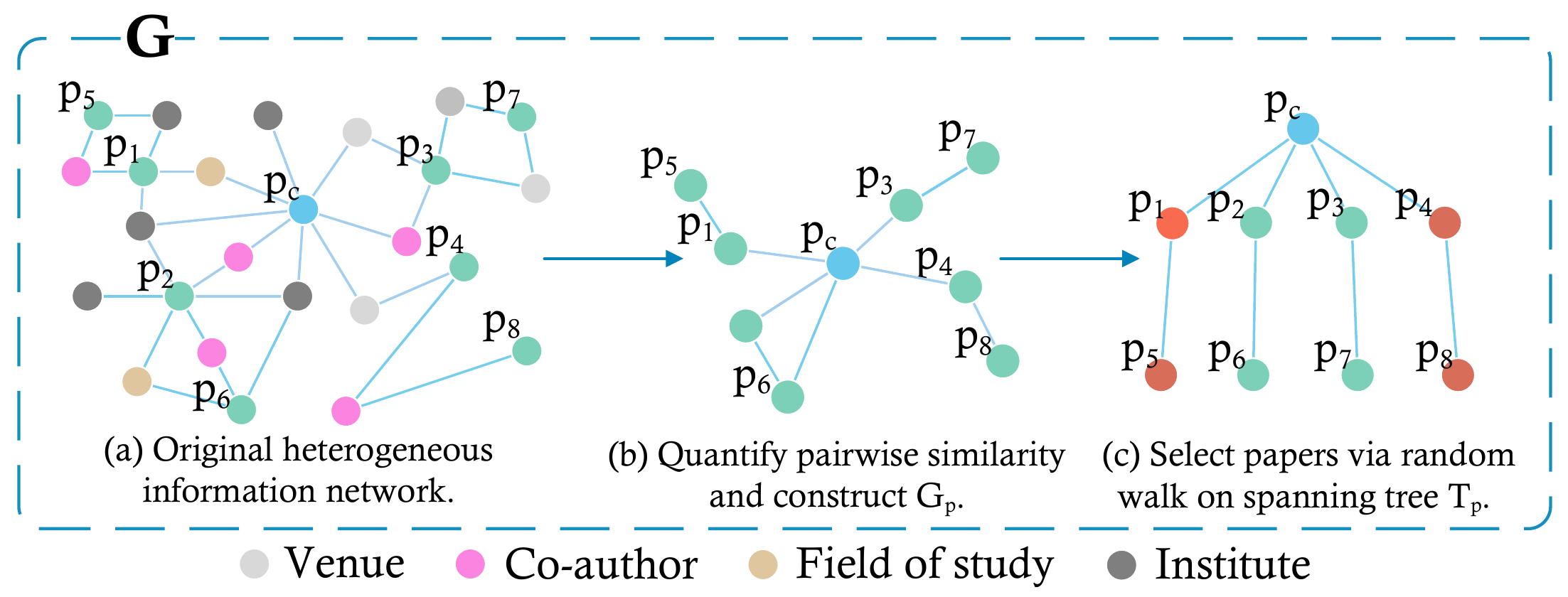}
\caption{Generating strategy of the generative module.}
\label{fig:G_framework}
\end{figure}

\subsubsection{Implementation of Generative Module} 
The generator aims to select the papers which are possibly homogeneous from the constructed HIN.
Once the discriminator cannot distinguish whether the papers are selected by the generator, the generator is guided to find the rules to select the homogeneous papers.
To update $\theta_G$, we follow \cite{policy1} to compute the gradient of $V(G, D)$ by policy gradient:
\begin{equation}
\begin{split}
&\nabla_{\theta_G} V(G, D) \\
& =  \sum_{p_i \in P}\mathbb{E}_{p \sim G(\cdot|p_k)}[\nabla_{\theta_G}\log G(p|p_k) \log (1-D(p, p_k))].\\
\end{split}
\label{eq:Ggradient}
\end{equation}

During each $G$ iteration, the generator selects the most similar papers from the HIN. 
The reward $\log (1-D(p, p_k))$ from the discriminator pushes the generator to update $\theta_G$, thus the similarity among papers will finally indicate the homogeneity among papers.

As for the quantification of similarity, a straightforward way is to define it as a softmax function over all other papers: 
\begin{equation}
G(p|p_k)=\frac{\exp(\vec{g}_{p}^T\vec{g}_{p_k})}{\sum_{p \in P, p\neq p_k}\exp(\vec{g}_{p}^T\vec{g}_{p_k})},
\label{eq:g_softmax}
\vspace*{1pt}
\end{equation}
where $\vec{g}_{p}$, $\vec{g}_{p_k}$ are the k-dimension representation vectors of papers $p$ and $p_k$ respectively for generator. 
And the parameters $\theta_G$ are the union of all $\vec{g}$ vectors. 

However, two limitations still exist: 
\begin{enumerate}
\item It entirely depends on the reward from the discriminator, ignoring the content information and relation information. 
We expect a way that generator can comprehend the information and make a wiser selection.

\item  It is time-consuming, because the similarity between each pair of papers need to be calculated. A more efficient generating strategy is required for the large-scale application.

\end{enumerate}

Here, we describe an information-aware generating strategy in detail, which is shown in Figure~\ref{fig:G_framework}.

At first, let $\mathcal{N}_p^{(i)}$ be the set of the papers that have $i$-order relation with $p$. 
For $i=1$, we define the homogeneity probability of $p \in \mathcal{N}_p^{(1)}$ given $p_k$ as follows:
\begin{equation}
\text{Pr}(p|p_k) = \frac{\sum_{t_m \in T_{share}}\exp(\vec{g}_{p}\vec{g}_{t_m}^T\cdot 
\vec{g}_{p_k}\vec{g}_{t_m}^T)}{\sum_{p_j \in \mathcal{N}_{p_k}^{(1)}}\sum_{t \in T} 
\exp(\vec{g}_{p_j}\vec{g}_{t}^T\cdot \vec{g}_{p_k}\vec{g}_{t}^T)},
\label{eq:pairwise}
\end{equation}
where $g_{p_i}$ is the representation of papers for $G$, and $g_{t_m}$ is the representation of $t \in V \backslash P$. 
It indicates that the papers that are connected by more entities are more possibly written by the same author. 

We then define $G(p|p_k)$ as follows:
{\small
\begin{align}
\centering
G(p|p_k) = \begin{cases} \text{Pr}(p|p_k), & \mbox{if } p \in \mbox{ $\mathcal{N}_{p_k}^{(1)}$}; \label{eq:G}
\\ \sum_m\text{Pr}(p|p_m)G(p_m|p_k), & \mbox{if } p \in \mbox{ $\mathcal{N}_{p_k}^{(i)}$, $i\neq 1$}.
\end{cases}
\end{align}}

Eq.~\eqref{eq:G} models the possibility of homogeneity among papers which have high-order connections. 
In practice, two papers written by the same author can be connected by a complicated path instead of two edges, e.g., $p-A-p_i-I-p_k$, $p$ and $p_k$ do not have straight connection, but they both have close relation with $p_i$, indicating that all of three papers are written by the same author. 

Since Eq.~\eqref{eq:G} is computationally inefficient, we implement it with the help of paper network. 
Based on the heterogeneous network, we construct the paper network $G_p$ at first. The weights of edges are decided by Eq.~\eqref{eq:pairwise}. Then we construct a tree $T_{p_k}$: (i) Add the given paper $p_k$ into $T_{p_k}$; (ii) Add the edge $(p_i, p_j)$ with highest weight into $T_{p_k}$, where $(p_i, p_j) \notin T_{p_k}$; (iii) Repeat step 2 until all papers in $G_p$ are added into $T_p$. 
Then there is a path $P_{p_k\rightarrow p_i}$ from $p_k$ to $p_i$ on the spanning tree $T_p$. 
The $G(p|p_k)$ is simplified as follows:
\begin{equation}
\centering
G(p|p_k)=\Pi_{(p_{m}, p_{m+1}) \in P_{p_k\rightarrow p_i}}\text{Pr}(p_{m}|p_{m+1}).
\label{eq:final_G}
\end{equation}

A straightforward interpretation of this process based on spanning tree is that given a paper  $p_k$, we first select the papers that are very similar to it as the homogeneous group. 
Then papers that are similar to the papers in the group are also possibly written by the same author. 
The spanning tree based strategy preserves the high-order connections among papers, which integrates the relation information.

Next, we discuss the selecting strategy for generator. We perform a random walk on $T_p$ starting at paper $p_k$ with respect to Eq.~\eqref{eq:pairwise}. During this process, once the generator decides to visit a paper that has been visited, the random walk is halted and the papers in the path will be selected by generator. These papers are selected by generator as the homogeneous papers with $p_k$, then they will be fed into discriminator as negative papers.

The algorithm maintains time efficiency and information-awareness:
\begin{itemize}
\item Given a paper, the generator only considers papers from the connected component $\mathcal{N}_{p_k}$ as candidates, which means there is no need to calculate the pairwise possibility with all the other papers. 
\item While selecting papers, it takes the information from the heterogeneous network into consideration. First, to calculate the pairwise possibility, Eq.~\eqref{eq:pairwise} integrates the relation information from the heterogeneous network. Besides, the random walk based generating algorithm comprehensively takes the high-order connection among papers into consideration.
\end{itemize}

\subsection{Clustering}
Based on the final representation $\vec{d}$, $\vec{g}$ of papers, we perform hierarchy agglomerative clustering (HAC) to partition $N$ papers into disjoint homogeneous sets. 

The process for the author name disambiguation is summarized in Algorithm \ref{al:all}.

\begin{algorithm}[h]
\small
\caption{\textbf{The proposed framework.}}\label{al:all}
\KwData{Paper set $P^a$ and information set $I^a$, $R^a$.}
\KwResult{The partition result, $\Phi(P^a|I^a, R^a) \to C^a$.}
Construct the heterogeneous informaton network $G^a$\;
Utilize content2vec module to learn content representation $\vec{u}$\;
Utilize node2vec module to learn content representation $\vec{v}$\;
Initialize $G(p|p_k; \theta_G)$ and $D(p, p_k; \theta_D)$ based on $\vec{u}$, $\vec{v}$\;
\While{model not converge}{
	Construct $G_{p_k}$ and $T_{p_k}$ according to Eq.~\eqref{eq:pairwise}\;
	\For{G-steps}{
    	Perform random walk on $T_{p_k}$ and generate papers into $S_{generated}$ for each $p_k \in P^a$\;
        Update $\theta_G$ according to Eq. ~\eqref{eq:Ggradient}, ~\eqref{eq:pairwise}, ~\eqref{eq:final_G}\;
    }
    Sort papers based on $\vec{d}$ and select top-relevant papers into $S_{pseudo}$ for $p_k\in P^a$\;
    \For{D-steps}{
    	Sample positive papers from $S_{pseudo}$ and negative papers from $S_{generated}$\;
        Update $\theta_D$ according to Eq. ~\eqref{eq:d},
        ~\eqref{eq:Dobjective}, ~\eqref{eq:Dgradient}\;
        }
}
Perform HAC algorithm based on representation result $\theta_D$ and $\theta_G$\;
\Return{$\Phi(P^a|I^a, R^a) \to C^a$\;}

\end{algorithm}

\begin{table*}[t]
\small
\centering
\caption{The detailed results on AceKG-AND.}
\vspace*{5pt}
\label{tb:detailed}
\scalebox{0.98}{
\begin{tabular}{l|lll|lll|lll|lll}
\toprule[0.1em]
 & \multicolumn{3}{c|}{Ours} & \multicolumn{3}{c|}{AMiner \cite{JieTang}} & \multicolumn{3}{c|}{\citet{Block}} & \multicolumn{3}{c}{\citet{Anonymized}} \\ \cline{2-13} 
Name & Prec & Rec & F1 & Prec & Rec & F1 & Prec & Rec & F1 & Prec & Rec & F1 \\ \hline
A. Kumar & 74.56 & 50.30 & 60.07 & 63.59 & 66.61 & \textbf{65.07} & 73.70 & 43.83 & 54.97 & 46.01 & 25.05 & 32.44 \\
Bo Jiang & 90.11 & 51.79 & 65.77 & 69.28 & 54.77 & 61.18 & 62.28 & 56.81 & 59.42 & 97.47 & 92.94 & \textbf{95.15} \\
Chi Zhang & 81.36 & 78.39 & \textbf{79.85} & 53.88 & 49.46 & 51.58 & 61.71 & 48.66 & 54.42 & 78.63 & 73.81 & 76.14 \\
Dong Xu & 95.64 & 93.38 & \textbf{94.50} & 78.46 & 73.61 & 75.96 & 39.72 & 100.00 & 56.86 & 96.12 & 62.64 & 75.85 \\
Fan Zhang & 80.37 & 80.07 & 80.22 & 50.76 & 66.67 & 57.64 & 72.80 & 84.60 & 78.26 & 92.95 & 80.87 & \textbf{86.49} \\
Hui Li & 65.83 & 43.90 & \textbf{52.68} & 56.53 & 32.98 & 41.65 & 59.60 & 36.32 & 45.14 & 58.26 & 24.45 & 34.45 \\
Jie Liu & 66.32 & 80.27 & \textbf{72.63} & 47.94 & 28.93 & 36.08 & 47.80 & 38.45 & 42.61 & 84.39 & 49.32 & 62.26 \\
Jie Yang & 91.39 & 88.72 & \textbf{90.03} & 71.60 & 70.68 & 71.14 & 46.90 & 55.61 & 50.89 & 90.18 & 72.34 & 80.28 \\
Lin Ma & 92.82 & 85.33 & \textbf{88.92} & 60.55 & 63.46 & 61.97 & 66.35 & 65.51 & 65.93 & 87.73 & 68.42 & 76.88 \\
Lin Zhang & 76.13 & 73.62 & \textbf{74.85} & 70.20 & 55.80 & 62.18 & 53.05 & 38.69 & 44.74 & 91.18 & 59.38 & 71.93 \\
Qian Wang & 96.55 & 84.70 & \textbf{90.24} & 73.80 & 73.02 & 73.40 & 70.19 & 63.96 & 66.93 & 85.04 & 74.71 & 79.54 \\
Tao Chen & 89.50 & 82.23 & \textbf{85.71} & 63.86 & 40.28 & 49.40 & 53.40 & 44.31 & 48.43 & 90.91 & 41.80 & 57.27 \\
Wei Gao & 92.62 & 94.99 & \textbf{93.79} & 78.34 & 73.78 & 75.99 & 70.18 & 40.68 & 51.51 & 85.19 & 63.50 & 72.76 \\
Wei Lu & 71.60 & 55.90 & \textbf{62.79} & 53.88 & 45.01 & 49.04 & 52.45 & 34.11 & 41.34 & 62.44 & 31.19 & 41.60 \\
Yong Xu & 91.64 & 89.01 & \textbf{90.31} & 49.28 & 55.59 & 52.24 & 56.72 & 54.80 & 55.74 & 68.40 & 54.55 & 60.69 \\
 \bottomrule[0.1em]
\end{tabular}
 }

\end{table*}

\section{Experiment}
\subsection{Datasets}
To evaluate the proposed method, we collect two real-world author name disambiguation datasets for experiments:
\begin{itemize}
\item AMiner-AND\footnote{\small \url{https://www.aminer.cn/na-data}}. 
The dataset is released by \cite{JieTang}, which contains 500 author names for training and 100 author names for testing. We construct the heterogeneous network including papers, co-authors, author affiliations (which are referred as institutes in our model), keywords (which are referred as fields of study in our model) and venues. However, there is no abstract in this dataset, so we can only use the titles as our content information in the experiment on this dataset.  
To illustrate our model's ability to combine content information and relation information and to support the researches which study the author name disambiguation task using content information, we construct a new dataset collected from AceKG \cite{AceKG}. The benchmark dataset consists of 130,655 papers from 17,816 distinguished authors. Each sample has the relation information and content information required by the proposed model. The labeling process is carried out comprehensively based on the e-mail address of authors, the co-author information and the institute information. 
\end{itemize}

\begin{table}[t]
\begin{center}
\vspace{0pt}
\caption{Results of author name disambiguation.}
\label{tb:allresult}
\footnotesize
\scalebox{0.91}{
\begin{tabular}{l|ccc|ccc}
\toprule[0.1em]
& \multicolumn{3}{c|}{AMiner-AND} & \multicolumn{3}{c}{AceKG-AND}\\
\hline
Model & Prec & Rec & F1& Prec & Rec & F1 \\
\hline
Zhang and Al& 70.63 & 59.53 & 62.81 & 72.35 &54.24 & 60.71 \\
Louppe et al. & 57.09& \textbf{77.22} & 63.10 & 56.69 & 57.82 & 55.88 \\
AMiner  & 77.96 & 63.03 & 67.79 & 58.57 & 55.41 & 56.21 \\
Ours & \textbf{82.23} &67.23 & $\textbf{72.92}^*$ & \textbf{78.26} & \textbf{70.73}& $\textbf{73.71}^*$\\
\bottomrule[0.1em]
\end{tabular}}
\begin{tablenotes}
\small \item[*] $*$ indicates that the $F1$ score of our model is the significant result over other models, with $p$-value less than $10^{-6}$.
\end{tablenotes}
\end{center}
\end{table}

\subsection{Baselines}
We compare our model against three state-of-the-art name disambiguation methods. We perform the hierarchical agglomerative clustering algorithm based on the results from these models and compare them by the pairwise Precision, Recall, and F1-score.

\textbf{\citet{Anonymized}}: This model constructs three networks under each name reference. The vertices are authors and papers. The weights of edges represent the connections among them. A designed network embedding is learned with an aim to preserve the connectivity of the constructed networks.

\textbf{\citet{Block}}: This model trains a function to measure the similarity between each pair of papers using the carefully designed pairwise features, including author names, titles, institute names etc.

\textbf{AMiner~\cite{JieTang}}: This model designs a supervised global stage to fine-tune the word2vec result, and designs an unsupervised local stage based on the first stage. In the local stage, it constructs a paper network, where the weight of edge reflects the similarity among papers. Then it uses graph convolutional network to preserve the connectivity of the paper network and learn the representation of papers.

To further evaluate the performance of each module, we also compare our performance at different stages. 

\textbf{Con.} This is the result based on the content representation result produced by Doc2vec module. This module represents the abstract and title information by a vector.

\textbf{Rel.} This is the result based on the relation representation results, which maps the nodes in heterogeneous information network into low-dimension representation space.

\textbf{Dis.} The result is from discriminator which aims to distinguish whether two papers are homogeneous based on information representation and relation representation. 

\textbf{Gen.} The result is from the generator which aims to approximate the underlying homogeneity distribution and to extract the high-order connections on HIN.


\begin{figure*}[t]
		\begin{subfigure}{.24\textwidth}
			\centering
			\includegraphics[width=\textwidth]{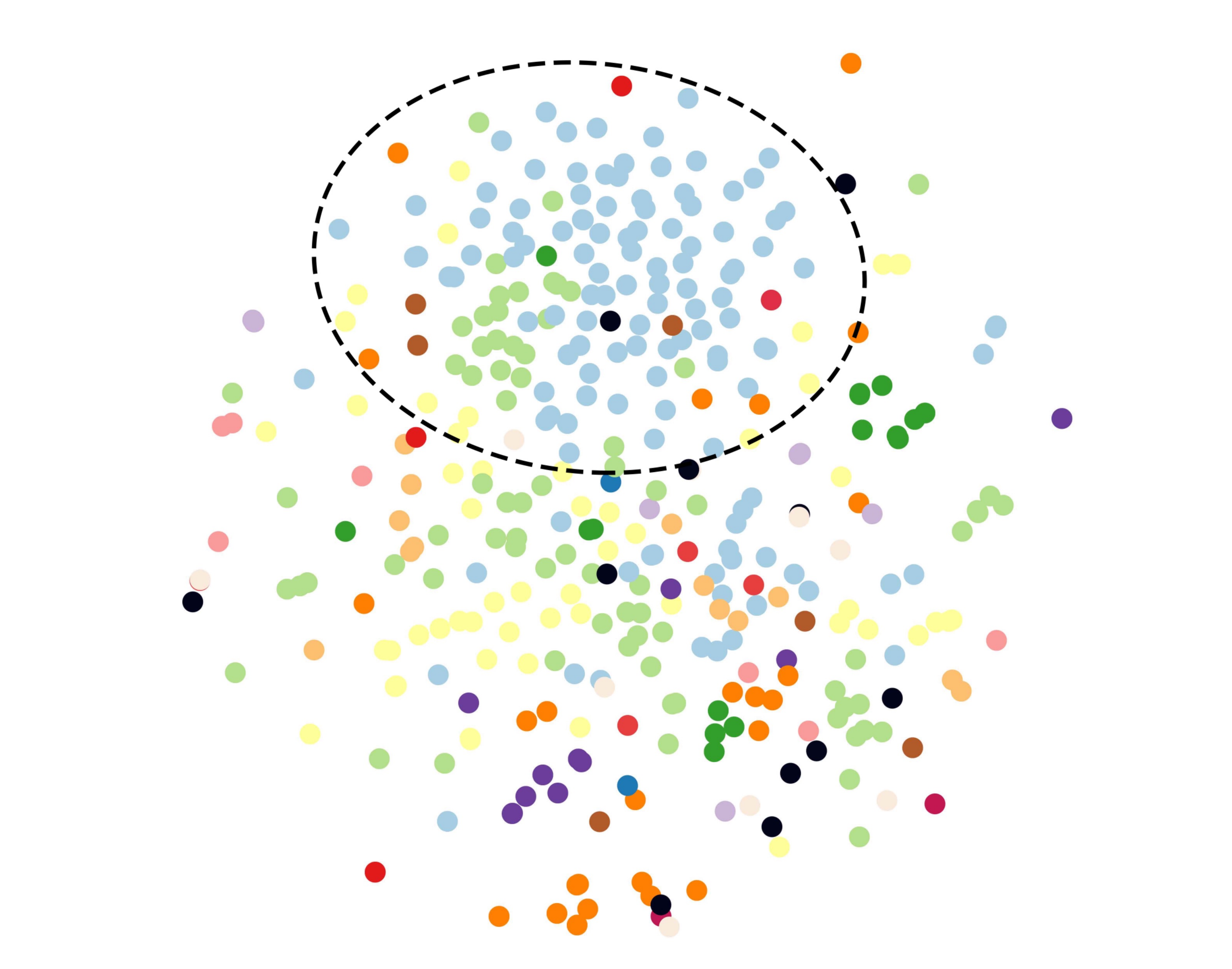}
			\caption{Con Emb.}
			\label{fig:subfig:a}
		\end{subfigure}
		\begin{subfigure}{.24\textwidth}
			\centering
			\includegraphics[width=\textwidth]{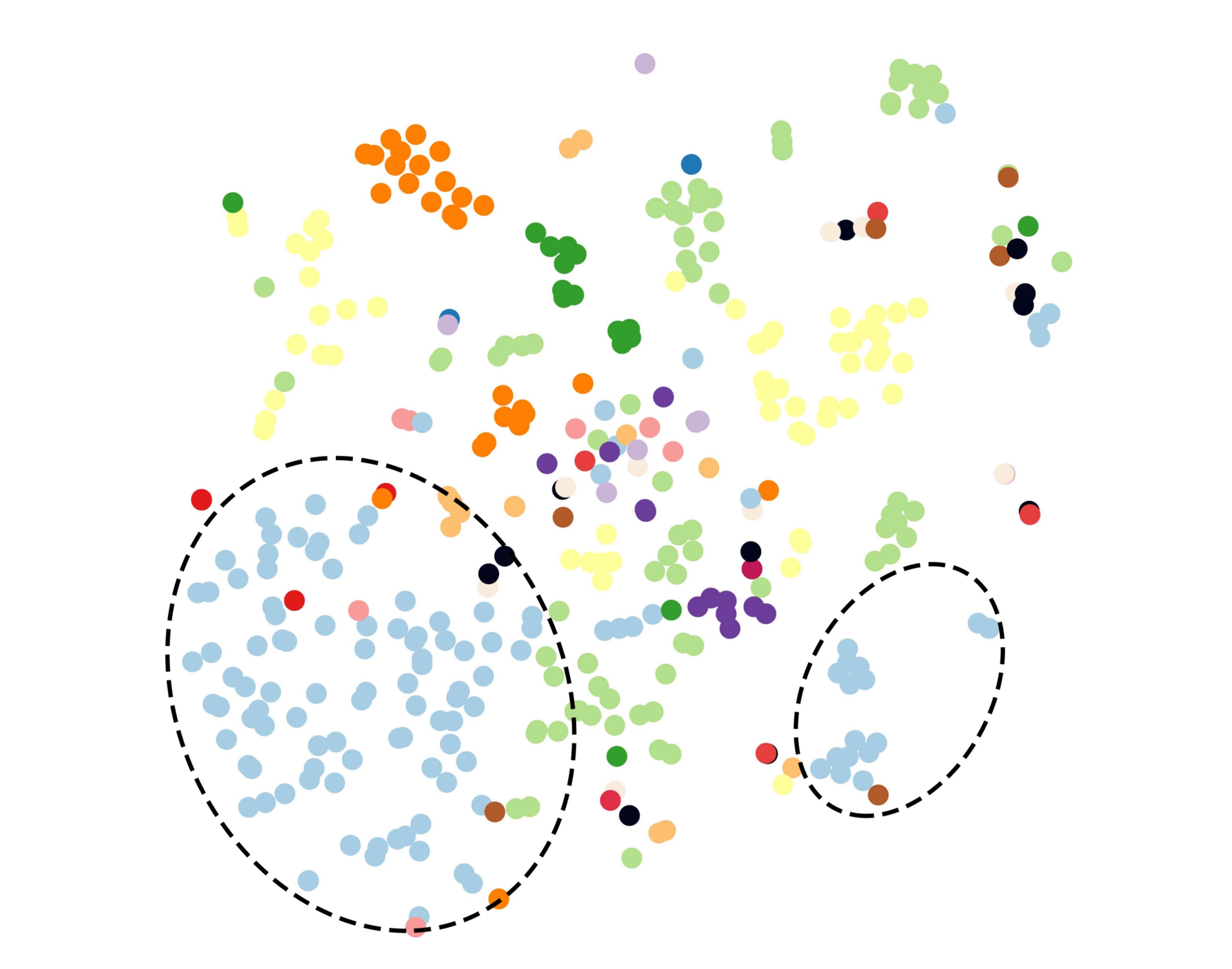}
			\caption{Rel Emb.}
			\label{fig:subfig:b}
		\end{subfigure}
		\begin{subfigure}{.24\textwidth}
			\centering
			\includegraphics[width=\textwidth]{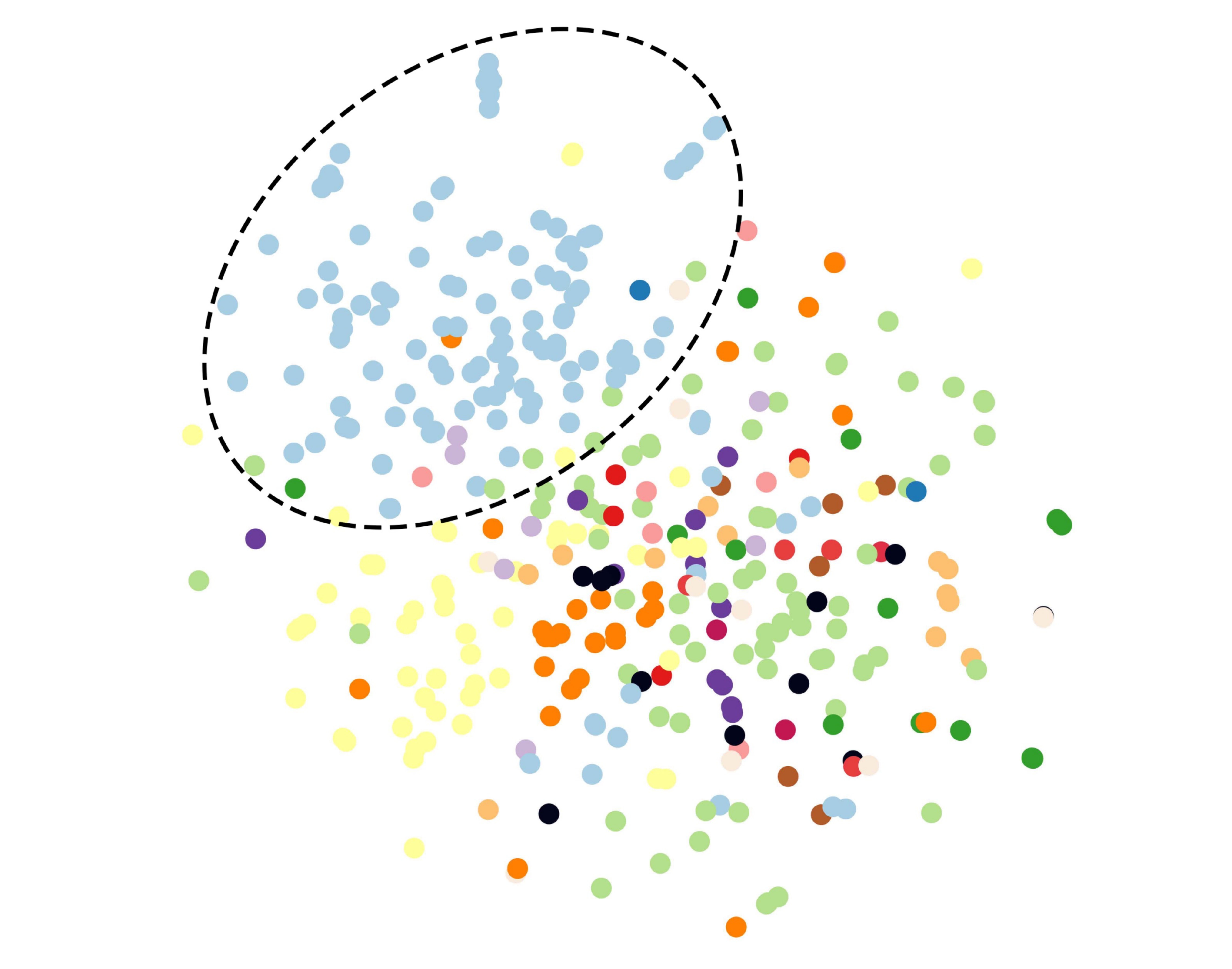}
			\caption{Dis Emb.}
			\label{fig:subfig:c}
		\end{subfigure}
		\begin{subfigure}{.24\textwidth}
			\centering
			\includegraphics[width=\textwidth]{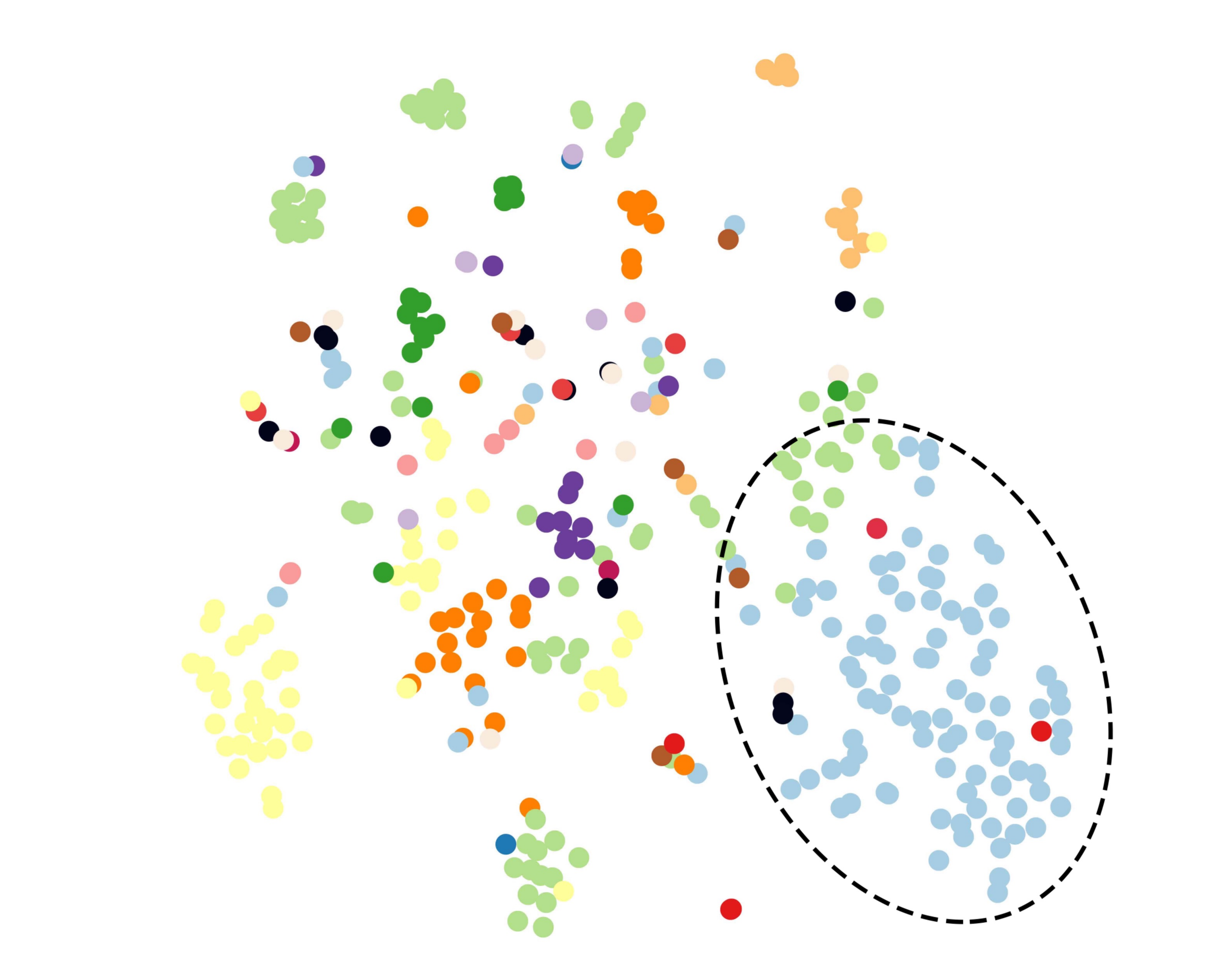}
			\caption{Gen Emb.}
			\label{fig:subfig:d}
		\end{subfigure}
		\vspace*{5pt}
		\caption{t-SNE Visualization of embedding spaces on a name reference \texttt{Yang Liu} in AceKG-AND. Each color in (a), (b), (c), (d) denotes a homogeneous cluster according to ground truth. Con Emb. represents the content representation result by Doc2vec. Rel Emb. represents the relation representation result by Node2vec. Dis Emb. and Gen Emb. are the results based on discriminator and generator results respectively. The dashed black ellipses circle the points of the same author.}
		\label{fig:2D}
	\end{figure*}
	
\subsection{Experiment Results}

We examine our model with several state-of-the-art models on AMiner-AND and AceKG-AND. In the experiment on AMiner-AND, we use 100 names for testing and compare the result with the results of other models reported in \cite{JieTang}. In the experiment on AceKG-AND, we sample 85 names for testing. Since \citeauthor{Block} and AMiner are supervised algorithms, the results from 5-folds cross-validation are reported. Hierarchy agglomerative clustering is performed on the results produced, where the number of clusters is given in advance.

Table \ref{tb:allresult} shows the overall performances of different models on two datasets. 
All the reported metrics are the macro-averaged scores of each metric of all test names. 
Our model outperforms all the other baselines by at least 5.13\% and 13.00\% in F1 score on the two datasets respectively.
On AMiner-AND, our model outperforms the baselines in terms of F1-score (+10.11\% over Zhang et al., +9.82\% over \citeauthor{Block} and +5.13\% over AMiner relatively).
On  AceKG-AND, the superiority is the same. 
As shown in Table~\ref{tb:detailed}, almost all the metrics of 15 random name references are improved by our model, which
demonstrates the significant superiority of our proposed method.

\begin{table}[t]
\caption{Results of components in the framework.}
\label{tb:contribution}
\begin{center}
\small
\begin{tabular}{l|ccc|ccc}
\toprule[0.1em]
& \multicolumn{3}{c|}{AMiner-AND} & \multicolumn{3}{c}{AceKG-AND}\\
\hline
Model & Prec & Rec & F1& Prec & Rec & F1 \\
\hline
Con & 15.74 & 9.31 & 11.30 & 69.57 & 47.68 & 55.40 \\
Rel & 74.32	& 51.38	& 56.34 & 69.74 & 45.84 & 53.65\\
Dis & \textbf{84.58} & 59.83 & 68.00 & \textbf{84.80} & 55.09 & 65.41\\
Gen & 82.23 & \textbf{67.23} & \textbf{72.92} & 78.26 & \textbf{70.73} & \textbf{73.71}\\
\bottomrule[0.1em]
\end{tabular}
\end{center}
\end{table}

\subsection{Ablation Analysis}
To evaluate the performance of each module, we also present our performance at different stages in Table \ref{tb:contribution}. 
It can be seen that the generative module achieves the most significant result. 
It can mine some high-order connections among papers and thus covers more homogeneous papers as candidate set.

Content representation module achieves a good result on AceKG-AND, 
while the result on AMiner-AND is low. Because this dataset only provides title as content information.
The experiment has illustrated that content information like abstract is valuable for this task.

The discriminative module achieves the highest Prec on two datasets. Because it mainly measures the pairwise similarity, the papers written by the same author
 can be discovered precisely. 
The problem is that it solves the problem from a local perspective, which leads to a low Rec result.

Experiments show that relation representation results achieve F1-scores 56.34\% and 53.65\% on two datasets respectively. 
For those homogeneous papers which are connected tightly by the relations, they are close in the relation representation space, which works for the clustering stage. However, for those which are content related but have few relations, this module can not group them together. 

\subsection{Embedding Analysis}
To dig into how each module works, we visualize the results of each stage in a 2-D way, which is presented in Figure \ref{fig:2D}. 
We analyze the layout of blue points in feature space. 
After a global measurement by content representation module, the papers in the same $C^a$ are preliminarily clustered together in Figure \ref{fig:subfig:a}. 
Figure \ref{fig:subfig:b} shows the results of relation representation module. It can be seen that homogeneous papers are grouped much better. 
The clustering results of discriminator and generator are much better, for they consider both of the content information and relation information. 
The blue points are grouped into one cluster successfully. 
And clusters in Figure \ref{fig:subfig:d} have clearer boundary than clusters in Figure \ref{fig:subfig:c}, which corresponds to the fact that the generator achieves a better result than discriminator.

\section{Conclusion}
In this paper, we propose a novel adversarial representation learning model for heterogeneous information network in the academic domain. 
We employ this model to deal with author name disambiguation task, which integrates the advantages from both generative methods and discriminative methods. 
To eliminate the requirement for labeled samples and to measure high-order connections among papers well, a self-training strategy for discriminator and a random walk based exploration for the generator are designed. 
Experimental results on AceKG-AND and AMiner-AND datasets verify the advantages of our method over state-of-the-art name disambiguation methods.
Besides, we plan to employ the proposed adversarial representation learning model on paper recommendation and mentor recommendation.

\section{ACKNOWLEDGMENT}
This work was supported by National Key R\&D Program of China 2018YFB1004700, NSF China under Grant 61822206, Grant 61960206002, Grant 61532012, Grant 61602303, Grant 61829201, Grant 61702327, Grant 61632017.

\bibliographystyle{aaai}
\bibliography{AAAI_WangH_7426}
\end{document}